\documentclass[twocolumn,superscriptaddress,nofootinbib,preprintnumbers,amsmath,amssymb]{revtex4}
\usepackage{latexsym}
\usepackage{graphicx}
\usepackage{amsthm}
\usepackage{float}
\usepackage{amsmath}
\usepackage{graphicx,subfigure}
\usepackage{fancybox,feynmp}
\usepackage{indentfirst}
\usepackage{epsfig}
\usepackage{epsfig,subfigure,psfrag}
\usepackage{dcolumn}
\usepackage{bm}
\usepackage{slashed}
\usepackage{cancel}
\usepackage{color}
\usepackage{tabularx}
\usepackage{hyperref}
\hypersetup{
    colorlinks=true,       
    linkcolor=blue,          
    citecolor=blue,        
    filecolor=blue,      
    urlcolor=blue           
}

\preprint{CERN-PH-TH-2015-023}

\newcommand{\be}{\begin{eqnarray}}
\newcommand{\ee}{\end{eqnarray}}
\newcommand{\beq}{\begin{eqnarray}}
\newcommand{\eeq}{\end{eqnarray}}

\newcommand{\bea}{\begin{eqnarray}}
\newcommand{\eea}{\end{eqnarray}}

\newcommand{\gev}{{~\rm GeV}}

\newcommand{\vev}[1]{{\langle #1 \rangle}}

\def\Tr{\mathop{\rm Tr}}
\newcommand{\Um}{\text{U}}
\newcommand{\SU}{\text{SU}}

\newcommand{\Eq}[1]{Eq.~(\ref{#1})}

\begin{document}

\title{Goldstone Gauginos}

\author{Daniele S. M. Alves}
\email{spier@nyu.edu}
\affiliation{Center for Cosmology and Particle Physics, Department of Physics, New York University, New York, NY 10003}
\affiliation{Department of Physics, Princeton University, Princeton, NJ 08544}

\author{Jamison Galloway}
\email{jamison.galloway@nyu.edu}
\affiliation{Center for Cosmology and Particle Physics, Department of Physics, New York University, New York, NY 10003}

\author{Matthew McCullough}
\email{matthew.mccullough@cern.ch}
\affiliation{Theory Division, CERN, 1211 Geneva 23, Switzerland}

\author{Neal Weiner}
\email{neal.weiner@nyu.edu}
\affiliation{Center for Cosmology and Particle Physics, Department of Physics, New York University, New York, NY 10003}

\date{\today}

\begin{abstract}
Models of supersymmetry with Dirac gauginos provide an attractive scenario for physics beyond the standard model. The ``supersoft'' radiative corrections and suppressed SUSY production at colliders provide for more natural theories and an understanding of why no new states have been seen. Unfortunately, these models are handicapped by a tachyon which is naturally present in existing models of Dirac gauginos. We argue that this tachyon is absent, with the phenomenological successes of the model preserved, if the right handed gaugino is a (pseudo-)Goldstone field of a spontaneously broken anomalous flavor symmetry.
\end{abstract}

\maketitle

\numberwithin{equation}{section}
\renewcommand\theequation{\arabic{section}.\arabic{equation}}
\renewcommand*{\thefootnote}{\fnsymbol{footnote}}
\section{\label{sec:intro} Introduction}
The results of LHC7 and LHC8 have directly challenged the idea that the weak scale is natural. The quadratic divergences of the Higgs mass associated with the top Yukawa, the Higgs' gauge interactions, and its self-interactions all should be cut off not too far above the Higgs mass. The top Yukawa contribution, in particular, should be cut off below the TeV scale for a ``natural'' theory with at most $\sim 10\%$ tuning. The possibility that colored partners may not be found at the LHC has pushed many to explore alternatives to a natural weak scale \cite{Wells:2003tf,ArkaniHamed:2004fb,Giudice:2004tc,Fox:2005yp,Arvanitaki:2012ps,ArkaniHamed:2012gw}.

As a general idea, there is nothing wrong with the fact that no colored partners have been found. A TeV mass top partner need not have been found yet, since a correction $\delta m_h^2~\approx~\frac{3}{8 \pi^2} {\rm TeV}^2~\sim~(200 \gev)^2$ would not signal an unnatural theory per se.
The problem, however, is that this understates the divergence in many cases. In supersymmetry, in particular, this quadratic divergence is softened only to a logarithmic divergence, with $\delta m_{H_u}^2~\approx~\frac{3}{4 \pi^2} m_{\rm stop}^2 \log(\Lambda/m_{\rm stop})$. If this divergence is cut off at $16 \pi^2 m_{\rm stop}$, the theory is generally tuned at the few percent level.

In typical supersymmetric theories, this tuning is aggravated by the gluinos. Gluinos, while not directly coupling to the Higgs, indirectly contribute to $\delta m_{H_u}^2$ through their radiative contributions to the soft masses of squarks. Simultaneously, $\tilde q \tilde g$ searches are often stronger than $\tilde q \tilde q$ searches alone, and $\tilde q \tilde q$ production is enhanced by a light gluino. 

Many of these issues can be ameliorated if gauginos are Dirac rather than Majorana. Dirac gauginos yield radiative corrections that are ``supersoft,'' in that they are cut off without any divergences. The squark mass is then naturally a factor of $\sim 4  - 5$ below the gluino mass, where the contributions to the Higgs soft masses are cut off,
and the logarithmic enhancement is minimal \cite{Fox:2002bu}. 

If a new SM adjoint superfield $A^i$ is introduced, a Dirac gaugino state with mass $m_D$ can be generated in the presence of a hidden $U(1)_B$\footnote{We label this $\Um(1)_B$ as it can be identified as a baryon number in SUSY QCD models where Goldstone Gauginos can be simply realized.} $D$-term expectation value, $\vev{W^B_\alpha} = \theta_\alpha D$, by the operator
\be
W_{\rm CS}=\frac{g}{M} W_B^\alpha W_\alpha^i A^i,
\label{eq:ssclassic}
\ee
with $W^i_\alpha$ the field strength of a SM gauge group with gauge coupling $g$.
This operator then marries the SM gaugino to the fermionic component of $A^i$ \cite{Fayet:1978qc,Fox:2002bu}. Such an operator preserves an $R$-symmetry which forbids a Majorana gaugino mass, and generates a mass for the real scalar in $A^i$ while leaving its imaginary component massless.

With purely Dirac gauginos, there are no $t$-channel gluino induced processes such as $qq \rightarrow \tilde q \tilde q$ at the LHC, making such theories ``supersafe'' from experimental constraints \cite{Choi:2008pi,Kribs:2012gx}.

There are problems with this setup, however. Unification is constrained to be within $SU(3)^3$, as a full adjoint of $SU(5)$ contributes as five flavors \cite{Fox:2002bu}, inducing a Landau pole for QCD well below the GUT scale \cite{Moroi:1993zj,Brahmachari:1993zi}. This is merely aesthetic, however. More pressing is the so-called ``lemon-twist'' problem. In addition to the ``classic'' supersoft operator of Eq.~(\ref{eq:ssclassic}), hereafter CS, a second supersoft operator is allowed
\be
W_{\rm LTS} = \frac{1}{M^2} W_B^\alpha W_{B\alpha} A^i A^i,
\label{eq:lt}
\ee
which has been termed the ``lemon-twist'' supersoft operator, hereafter LTS (see e.g., \cite{Carpenter:2010as}). This operator generates a $B\mu$-type term for the scalar adjoints, destabilizing one direction of the adjoint and thereby breaking SM symmetries. If both operators of Eqs.~(\ref{eq:ssclassic}, \ref{eq:lt}) are generated at one loop, we encounter a standard $\mu$-$B\mu$ problem where the tachyon is not only present, but cannot be removed by the contribution to the scalar mass from the classic supersoft operator, which is $\mathcal O(4 \pi/g)$ smaller than the contribution from LTS \cite{Fayet:1978qc,Fox:2002bu,Benakli:2008pg,Benakli:2010gi,Arvanitaki:2013yja,Csaki:2013fla}. 
One can cancel this operator with additional soft scalar masses, which may be generated with $D$-term spurions \cite{Nelson:2015cea}, but typically this comes at the cost of nullifying the theory's supersoftness. Alternatively, one can include an explicit Majorana mass term, $\Delta W = \mu A^2$ with $\mu \sim (4 \pi /g) m_D$, but this would render the low energy gluino Majorana, with a mass $m_D^2/\mu \sim \alpha\,\mu/4 \pi$.  This reintroduces a large logarithmic contribution to the Higgs soft masses, and removes supersafeness (without even beginning to worry about the origin of the scale for $\mu$).

This operator seems impossible to forbid \cite{Fox:2002bu}, as the presence of $W^\alpha W_\alpha$ along with $W^\alpha W'_\alpha A$ would seem to imply that no symmetry can preclude $( W'^\alpha A)^2$. 

In this Letter, we show that if the CS operator originates from the spontaneous breaking of an anomalous global symmetry, the right handed gaugino is the fermionic component of a Goldstone superfield, and LTS is not generated. We will argue that CS is connected to a mixed anomaly, while LTS is not, providing a natural framework within which Dirac gauginos are viable.

\section{\label{sec:ggaugino} A Goldstone Gaugino}
Let us begin by understanding the $\mu$-$B\mu$ problem for Dirac Gauginos in the context of a simple messenger model. We take a superpotential
\be
\label{eq:messengermodel}
W_{\rm UV} = \mu\,  \overline T\, T + \lambda\, \overline T A\, T,
\ee
and assume an $N$-plet of messengers ($T, \overline T$), which transforms nontrivially under the SM, and whose scalars are split by a $D$-term such that their masses are $\mu^2 \pm D$. Integrating the  messengers out, both CS and LTS are generated,
\begin{eqnarray}
W_{\rm IR} &=& \frac{\sqrt 2\lambda }{16 \pi^2}\frac{g}{\mu} \,W_B^\alpha W_\alpha^i A^i, \nonumber \\
&&+\, \frac{\lambda^2}{32 \pi^2} \frac{1}{\mu^2}\, W_B^\alpha W_{B\alpha} A^i A^i.
\end{eqnarray}
The LTS operator above contributes to the masses of the scalar and pseudoscalar adjoints as $\delta m_{a_R}^2$=$-(4\pi/\alpha)\,m_D^2\, ,\;\delta m_{a_I}^2$=$(4\pi/\alpha)\,m_D^2$, respectively, while the contribution from the CS operator is $\delta m_{a_R}^2$= $4\,m_D^2$, $\delta m_{a_I}^2$= $0$. In particular the scalar adjoint becomes a tachyon due to the dominant contribution from LTS.

Let us now introduce a novel realization of Dirac gauginos, which we will refer to as the ``Goldstone Gaugino" (GoGa) mechanism. We begin by promoting $A$ to a nonlinear sigma field,
\be
 A=f\,e^{\Pi^iT^i/f},  \: \: \Pi^i \, |_{\theta,\bar\theta=0} = s^i+i\pi^i,
 \ee
and setting $\mu=0$ in (\ref{eq:messengermodel}), so that we have
\be
\label{eq:GGmessengers}
W' =  \lambda f\; \overline T \,e^{\Pi/f}\, T,
\ee
where $\Pi \equiv \Pi^i T^i$ and $T^i$ are $\SU (N)$ generators. The messengers are now massive only as a result of spontaneous symmetry breaking, whose dynamics will be left unspecified for the moment.

There are a number of ways to see what is achieved with this mechanism. To begin, let us simply study this perturbatively.

The model in (\ref{eq:GGmessengers}) resembles the one in (\ref{eq:messengermodel}) up to higher order terms in $\Pi$, i.e.
\be
\label{eq:GGmessengersexpanded}
W' =  \lambda f\; \overline T \, T+ \lambda\; \overline T\, \Pi \,T + \frac{\lambda}{2f}\, \overline T\, \Pi^2 T +...
\ee
The third term becomes important when the messengers are integrated out, such that its contribution to the generation of LTS cancels that of the previous terms.  To see this more explicitly, consider the (bosonic part of the) potential for the pseudoscalar adjoint $a_I$ originating from ($\ref{eq:messengermodel}$):
\be
V^{\text{scalar}}_{a_I}= 2\lambda^2 \big(\,|t |^2 +|\bar t|^2\big) \,a_I^2.
\ee
When the scalars $t$ and $\bar t$ are integrated out, a mass is generated for $a_I$. For the GoGa model described in (\ref{eq:GGmessengersexpanded}), on the other hand, the presence of the additional superpotential term $\propto \overline T\, \Pi^2 T$ guarantees that pseudoscalar quartic couplings of the form $\propto (|t |^2 +|\bar t|^2)\pi^2$ vanish. In fact there are no tree-level scalar potential terms for the pseudoscalar $\pi$,
\be
V^{\text{scalar}}_\pi= 0,
\ee
and hence no mass is generated for $\pi$ when $t$, $\bar t$ are integrated out. 

We can also see this easily without any expansion:  a simple inspection of the superpotential in (\ref{eq:GGmessengers}) reveals that $|\partial W/\partial T|^2,|\partial W/\partial \bar T|^2,|\partial W/\partial \Pi|^2$ are all manifestly independent of $\pi$. The potential {\em does} depend on the real scalar, $s$,
and so the absence of LTS contributions to $m_s^2$ is less trivial to see. It depends on a cancellation of diagrams generated from $s(|t |^2 +|\bar t|^2)$ and $s^2(|t |^2 +|\bar t|^2)$ interactions. Nonetheless, as LTS generates masses for both scalar and pseudoscalar components satisfying $\delta m_{s}^2=-\delta m_{\pi}^2$, and since we can see that $\delta m_{\pi}^2=0$, $\delta m_{s}^2$ contributions from LTS necessarily vanish in the GoGa model of ($\ref{eq:GGmessengers}$). (The full, diagrammatic calculation can be found in the appendix of \cite{Alves:2015bba}.)

In contrast, the couplings of the fermions are unchanged -- {\it i.e.}, there are no relevant higher order terms in the expansion of the exponential of ($\ref{eq:GGmessengers}$) -- and thus the CS operator is still generated, with $A^i$ being replaced with $\Pi^i$ in (\ref{eq:ssclassic}).

Another perspective on the GoGa mechanism is that the LTS operator is formally a correction to the holomorphic gauge coupling of $U(1)_B$, (even if for us it is not gauged, we can in principle gauge it). It is generated only if the adjoints $\Pi$, when treated as background fields, correct the masses of the messengers at $\mathcal{O}(\Pi^2)$ \cite{Csaki:2013fla}, which is not the case if $\Pi$ is a Goldstone coupling to the messengers as in (\ref{eq:GGmessengers}).

A final, simple argument as to why such a scenario is safe from the LTS tachyon is as follows: there is an approximate shift symmetry of $\Pi$ which is explicitly broken by the gauge coupling $g$. Therefore, any operators that violate the shift symmetry can only be generated if  accompanied by powers of $g$, such that in the limit $g\rightarrow 0$ the shift symmetry is restored and such operators vanish. This is the case for CS, which is $\propto g$. In contrast, the LTS operator has no connection to the gauge coupling, and hence cannot be generated.

\subsection{Anomalies and Classic Supersoft}
While we can calculate the size of the CS operator in a given perturbative theory, we would also like to understand when the GoGa mechanism occurs more generally. For instance, strongly coupled models of symmetry breaking are common, but it is not immediately obvious whether CS is generated or not.

We can understand the origin of the classic supersoft operator in its connection to the UV gauge anomalies of the theory.
Setting $\mu \!\!=\!\! 0$ in the messenger superpotential (\ref{eq:messengermodel}), the approximate global symmetry is enlarged to $\SU (F)_L \times \SU (F)_R \times \Um (1)_B$, with $A$ now promoted to a bifundamental.  This global symmetry is only weakly broken by the gauging the vector subgroup  $\SU (F)_V$, which we identify with (all or part of) the SM gauge group.  As $A$ develops a VEV $\langle A \rangle = v \times {\bold 1}_{{}_{F}}$, the global symmetry is spontaneously broken to $\SU (F)_V \times \Um (1)_B$ and the messengers $T$,$\bar T$ can be integrated out. The remaining low energy degrees of freedom are given by the Goldstone superfields $\Pi^i$:
\be
\label{eq:GGparam}
A = (v + \sigma) \, e^{T^i \Pi^i / f} \:,\: \Pi^i \, |_{\theta,\bar\theta=0} = s^i+i\pi^i.
\ee

An essential feature of our mechanism is that the axial current $J_A^{\mu\, i}$ associated with the Goldstones $\pi^i$ is anomalous with respect to the gauging of $\SU (F)_V~ \times~\Um (1)_B$,
\be
\label{divJ}
\!\!\!\!\partial_\mu J^{\mu\, i}_A \!= \!\frac{1}{16\pi^2} B^{\mu\nu} \tilde{F}^{\, i}_{\mu\nu}\!+\!\frac{1}{48\pi^2} \Tr\big[ T^i  {F}^{\mu\nu} \tilde{F}_{\mu\nu} \big]
\ee
where $F^{\,i}_{\mu\nu}$ and $B^{\mu\nu}$ are the field strengths of $\SU (F)_V$ and $\Um (1)_B$, respectively.  Each term can be understood diagrammatically from the following:
\begin{center}
\includegraphics[width=6cm]{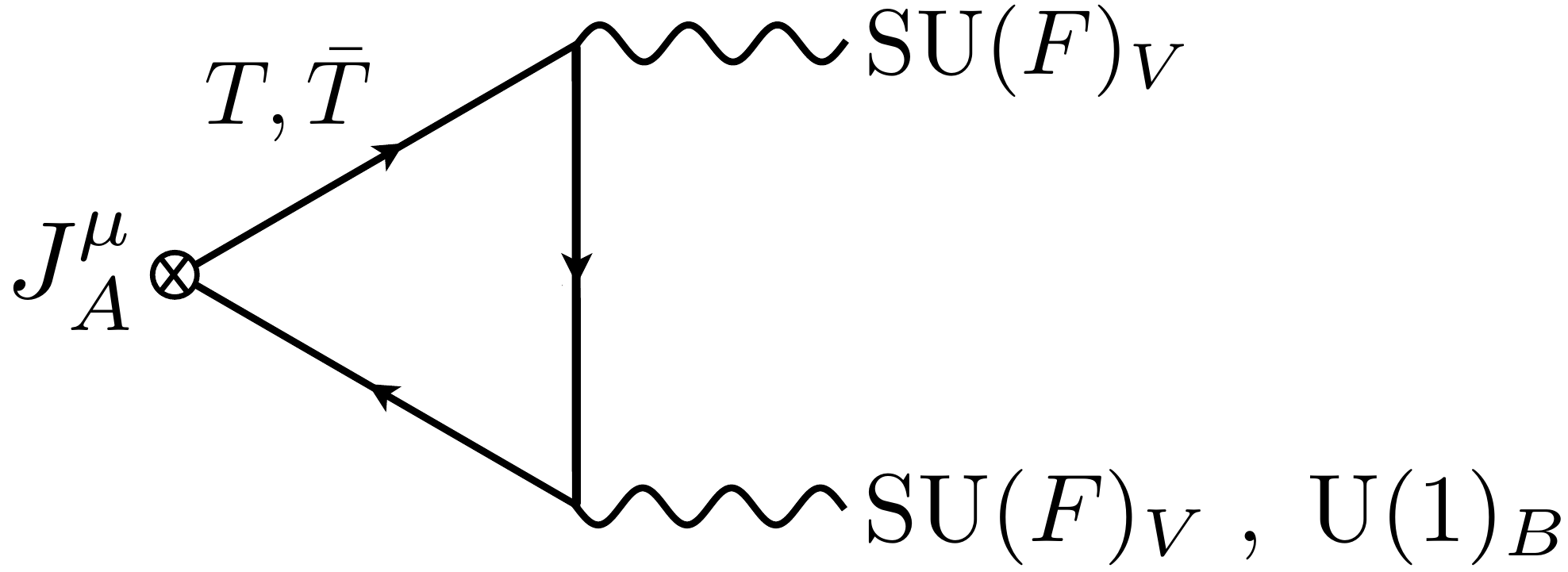}
\end{center}
This anomaly is accounted for in the low energy Lagrangian by the presence of a Wess-Zumino-Witten (WZW) term \cite{Wess:1971yu,Witten:1983tw}, whose bosonic part, to leading order in the $\pi^i$ fields, is given by
\be
\label{WZW}
\!\!\! \mathcal L \! \supset \! \frac{1}{ 16 \pi^2 f}\!\!\Tr\! \big[\pi  {B}_{\mu\nu} \tilde{F}^{\mu\nu}\big]\!+\!
 \frac{1}{48\pi^2 f} \!\!\Tr\! \big[ \pi   F_{\mu\nu} \tilde{F}^{\mu\nu} \big]
\ee
Interestingly, this supersymmetric chiral anomaly when expressed SUSY-covariantly in the superpotential becomes:
\be
\label{CS-WZW}
\!\!\!\! W\! \supset \! \frac{1}{16\pi^2 f} W_B^\alpha  W_\alpha^i  \Pi^i \! + \! \frac{1}{48\pi^2 f} \! \Tr \big[ W^\alpha  W_\alpha  \Pi \big]
\ee
and we can immediately recognize the first term as the CS operator (\ref{eq:ssclassic}).\footnote{In (\ref{divJ})-(\ref{CS-WZW}) the gauge coupling can be recovered by rescaling $A_\mu \rightarrow g A_\mu$.}

The LTS operator on the other hand is not related to gauge anomalies and does not stem from a WZW term.\footnote{This is familiar from ordinary QCD. The neutral pion couples to two $U(1)_{\rm EM}$ currents through a WZW term, while a coupling of {\it two} neutral pions to these same currents is absent when explicit breaking terms (e.g. quark masses) are taken to vanish.}
It is formally related to the holomorphic gauge coupling of $\Um (1)_B$, which depends only on $\text{log}(\text{det}A)$, and thus it is independent of the Goldstones $\Pi$.

Dangerous linear terms involving the singlet $S$ which marries the bino \cite{Fox:2002bu,Carpenter:2015mna}, namely $S\,W_B^\alpha W_{B \alpha}$, are not forbidden by a shift symmetry of the pseudoscalar component of $S$. Such a shift would simply generate a $\theta$-term for $\Um (1)_B$, which is a total derivative. Nonetheless, it is not hard to make this term absent, for instance by embeding hypercharge into a non-Abelian group, such that $S$ is associated with a traceless generator, or by imposing an $S\rightarrow -S$ parity \cite{Fox:2002bu}.

\subsection{A Perturbative model}
To realize the Goldstone Gaugino scenario, we must invoke dynamics that can spontaneously break a large flavor symmetry. 
This can be achieved by a simple symmetry breaking sector coupling $A$ to a singlet, $X$:
\be
W= \frac{X}{\Lambda^{F-2}} \big(\! \det A-v^F \big) .
\label{eq:non-ren}
\ee
In the vacuum the symmetry breaking structure is $\SU(F)_L \times \SU(F)_R \to \SU(F)_V$ as the $F$-term for $X$ requires a vacuum expectation value $\langle  A\, \rangle~=~v~\times~{\bf 1}_F$; the traceless components, $\Pi$, therefore become Goldstones of the spontaneously broken symmetry. While this superpotential is non-renormalizable, it captures the dynamics of certain SUSY QCD models which provide a UV completion, under which composite states of the microscopic quarks provide the right-handed gaugino \cite{Alves:2015bba}.

The Goldstone bosons associated with the broken generators can be parameterized as in Eq.~(\ref{eq:GGparam}).
Thus, if $A$ couples to messengers as in  \Eq{eq:messengermodel} with $\mu=0$, then this realizes the Goldstone Gaugino scenario. By inspection one can see that the arguments applied to Eq.~(\ref{eq:GGmessengers}) apply here as well, and no LTS operator is generated.

\section{Discussion}
It is fair to say at this point that much of the MSSM parameter space under consideration a decade ago has been excluded. This has come from a combination of direct and indirect searches, as well as from the discovery of the Higgs boson and its mass measurement. In light of this, alternatives that provide qualitatively different expectations are increasingly important.

Dirac gauginos are one such alternative. Because of their supersoft nature, they naturally provide a setup where heavy gluinos do not destabilize the weak scale, and allow heavier squarks with much smaller contributions to the Higgs mass. In addition, the present $R$-symmetry ameliorates a number of flavor and direct search constraints \cite{Kribs:2007ac,Frugiuele:2012pe}. They can also play a role in ``doubly invisible'' decay topologies in SUSY models, where existing limits for squarks are considerably weakened \cite{Alves:2013wra}.

Unfortunately, these models are plagued by a tachyonic instability which naturally arises in all simple models.  Previous solutions to this problem generally spoil supersoftness, yield a low energy theory with Majorana gauginos, or require extended messenger sectors with tuning of messenger mixings \cite{Amigo:2008rc, Csaki:2013fla}.

We have argued that there is a simple solution to this if the right handed gaugino is identified as a Goldstone field of some spontaneously broken anomalous global symmetry. In this Goldstone Gaugino scenario, the resulting shift symmetry is broken by the presence of the gauge interactions. Consequently, the classic supersoft operator is naturally generated, reflecting the anomalies of the theory, while the ``lemon-twist'' operator that produces the tachyon is not. This also provides a natural explanation why the right handed gaugino field does not have a high scale mass. 

The shift symmetry can arise from a spontaneously broken flavor symmetry in the effective theory, or may reflect a spontaneously broken symmetry at a higher energy scale and be realized nonlinearly. 

This basic GoGa framework provides a simple understanding of how a phenomenologically viable Dirac gaugino model can arise, without the problem of any tachyons. In light of this, phenomenologists and experimentalists should be free to consider Dirac Gaugino models as viable contenders to provide more natural models of the weak scale.

\begin{acknowledgements}
We thank Nathaniel Craig, Sergei Dubovsky, Claudia Frugiuele, Brian Henning, Anson Hook, Patrick Fox, Adam Martin, John Terning, and Jay Wacker for useful discussions. D.A. is supported by NSF-PHY-0969510 (the LHC Theory Initiative). N.W. and D.A. are supported  by the NSF under grants PHY-0947827 and PHY-1316753.  J.G. is supported by the James Arthur Postdoctoral Fellowship at NYU.

\end{acknowledgements}

\bibliographystyle{JHEP}
\bibliography{REFS}

\end{document}